\documentclass[preprint,superscriptaddress,nofootinbib,eqsecnum]{revtex4-1}
\usepackage[utf8]{inputenc}
\usepackage{amsmath,amssymb,amsthm}
\usepackage{graphicx}  
\usepackage[dvipsnames,x11names]{xcolor} 
\usepackage{float}
\usepackage{mathtools}
\usepackage{hyperref}

\newcommand{\beq}{\begin{equation}}
\newcommand{\eeq}{\end{equation}}

\DeclareMathOperator{\diag}{diag}

\DeclarePairedDelimiter\abs{\lvert}{\rvert}%
\DeclarePairedDelimiter\norm{\lVert}{\rVert}%

\makeatletter
\let\oldabs\abs
\def\abs{\@ifstar{\oldabs}{\oldabs*}}
\let\oldnorm\norm
\def\norm{\@ifstar{\oldnorm}{\oldnorm*}}
\makeatother

\begin{document}

\title{CKM substructure from the weak to the Planck scale}

\author{Yuval Grossman}
\email{yg73@cornell.edu}
\affiliation{Department of Physics, LEPP, Cornell University, Ithaca, NY 14853, USA}

\author{Ameen Ismail}
\email{ai279@cornell.edu}
\affiliation{Department of Physics, LEPP, Cornell University, Ithaca, NY 14853, USA}

\author{Joshua T. Ruderman}
\email{ruderman@nyu.edu}
\affiliation{Center for Cosmology and Particle Physics, Department of Physics, New York University, New York, NY 10003, USA}

\author{Tien-Hsueh Tsai}
\email{tt423@cornell.edu}
\affiliation{Department of Physics, LEPP, Cornell University, Ithaca, NY 14853, USA}
\affiliation{Department of Physics, National Tsing Hua University, Hsinchu 300, Taiwan}

\date{\today}

\begin{abstract}

We look for relations among CKM matrix elements that are not consequences of the Wolfenstein parametrization. In particular, we search for products of CKM elements raised to integer powers that approximately equal $1$. We study the running of the CKM matrix elements and resolve an apparent discrepancy in the literature. To a good approximation
only $A$ runs, among the Wolfenstein parameters. %, and we provide a quadratic fit to its running.
Using the Standard Model renormalization group we look for CKM relations at energy scales ranging from the electroweak scale to the Planck scale, and we find 19 such relations.
These relations could point to structure in the UV, or be numerical accidents.  For example, we find that $\abs{V_{td} V_{us}}=\abs{V_{cb}^2}$, within $2\%$ accuracy, in the $10^9$--$10^{15}$~GeV range. We discuss the implications of this CKM relation for a Yukawa texture in the UV.

\end{abstract}

\maketitle

%%%%%%%%%%%%%%%%%%%%%%%%%%%%%%%%%%%%%%%%%%%%%%%%%%%%%%%%%%%%%%%%%%%%%%%%%%%%%
\section{Introduction}\label{sec:intro}
%%%%%%%%%%%%%%%%%%%%%%%%%%%%%%%%%%%%%%%%%%%%%%%%%%%%%%%%%%%%%%%%%%%%%%%%%%%%%
The CKM flavor-mixing matrix possesses nontrivial structure, as exemplified by the Wolfenstein parametrization~\cite{Wolfenstein:1983yz}:
\begin{equation}\label{eq:wolfenstein}
    V = \begin{pmatrix}
        1 - \lambda^2/2 & \lambda & A \lambda^3 (\rho - i \eta) \\
        -\lambda & 1 - \lambda^2/2 & A \lambda^2 \\
        A \lambda^3(1 - \rho - i\eta) & -A \lambda^2 & 1
    \end{pmatrix}
    + \mathcal{O}(\lambda^4).
\end{equation}
Experimentally, $\lambda \approx 0.22$, reflecting the hierarchical nature of the CKM matrix. Eq.~\eqref{eq:wolfenstein} is an expansion in the small parameter $\lambda$, with the other Wolfenstein parameters $A$, $\rho$, and $\eta$ taken to be $\mathcal{O}(1)$. It is natural to wonder whether there is structure in the CKM matrix beyond the Wolfenstein parametrization. This could manifest as precise relations between elements of the CKM matrix (or equivalently, relations between the Wolfenstein parameters) that do not follow purely from the smallness of $\lambda$ or the unitarity of $V$.

To this end, Ref.~\cite{Grossman:2020qrp} introduced the concept of ``Wolfenstein anarchy'', in analogy with anarchic models of the PMNS lepton mixing matrix~\cite{Hall:1999sn,Haba:2000be,deGouvea:2003xe,deGouvea:2012ac,Lu:2014cla}. In contrast to the CKM matrix, it is unknown whether the PMNS matrix possesses any nontrivial strucure. PMNS anarchy is the concept that there is no such structure: all of the elements of the PMNS matrix are comparable and there are no new relations among them~\cite{Hall:1999sn}. Following this, a CKM matrix is Wolfenstein anarchic if it is generic other than the one small parameter $\lambda$, and no relations exist other than those already implied by the Wolfenstein parametrization. The opposite situation is that the CKM matrix has a substructure. Whether the CKM matrix is Wolfenstein anarchic or has a substructure depends upon the UV completion underlying the flavor structure observed in nature.

Ref.~\cite{Grossman:2020qrp} identified two novel CKM relations that are approximately satisfied: 
\beq \label{eq:two-rels}
\abs{V_{td}^2} = \abs{V_{cb}^3}, \qquad \abs{ V_{ub}^2 V_{us} } = \abs{V_{cb}^4}.
\eeq 
However, the CKM matrix runs~\cite{Balzereit:1998id,Juarezwysozka:2002kx}, and so the relations in Eq.~(\ref{eq:two-rels}) should be understood as holding at a low scale, below the top quark mass $m_t$. (Below $m_t$, the CKM matrix is essentially constant~\cite{Zyla:2020zbs}.) If some UV physics generates CKM substructure, this should be reflected in CKM relations that appear at the scale of the new physics. These relations may not hold in the IR due to the running of the elements of CKM matrix. For this reason, we seek to explore CKM relations at different scales, from the weak scale to the Planck scale.

What can we learn from such relations? The great hope is that they are due to some fundamental reason --- that is, that they serve as a hint of UV physics. Yet, such relations can be totally accidental. Our hopes in this paper are twofold. First we look for relations. Once we have found them, we ask what we can learn about any possible UV model that generates them. 

The running of the CKM matrix is model-dependent. In this paper we consider the Standard Model (SM) and one realization  
of the Minimal Supersymmetric Standard Model (MSSM)~\cite{Nilles:1983ge,Haber:1984rc}. 
While in the SM the numerical values of the parameters at the weak scale are known to a good accuracy, for the MSSM we have to choose various unknown parameters. Our choice is to take all superpartner masses to be equal at the weak scale and $\tan \beta \equiv v_d / v_u = 1$. 
This realization of the MSSM is experimentally excluded, because the LHC typically constrains superpartner masses to be heavier than the electroweak scale (see for example Refs.~\cite{ATLAS:2020syg,CMS:2019ybf,ATLAS:2020dsf,CMS:2021beq}).  Heavier superpartner masses introduce logarithmic threshold corrections to the CKM running which we do not include here for simplicity.  Our purpose with this toy MSSM example is to demonstrate that the running can change when going beyond the SM, leading to different relations.

In both the SM and the MSSM, to a good approximation, only the Wolfenstein parameter $A$ runs. The running of $A$ has been calculated before, but there is an apparent disagreement in the literature about its running in the SM: Ref.~\cite{Balzereit:1998id} reports an increase in $A$ of about $13\%$ from the weak scale to the GUT scale, while in Fig.~2 of  Ref.~\cite{Juarezwysozka:2002kx} $A$ increases by about 25\%. We resolve this discrepancy. As we explain in \S~\ref{sec:running}, we find that recomputing the running of $A$ using the methods of Ref.~\cite{Juarezwysozka:2002kx} gives a result which, in fact, agrees with Ref.~\cite{Balzereit:1998id}. Thus, we take the results of both Ref.~\cite{Balzereit:1998id} and Ref.~\cite{Juarezwysozka:2002kx} (except for their Fig.~2) to be correct.

Having the forms of running of the CKM matrix, we then search for CKM relations up to the Planck scale in the SM and the MSSM\@. We find one particularly intriguing relation, 
\begin{equation}
    \abs{V_{td} V_{us}} = \abs{V_{cb}^2} ,
\end{equation}
that holds in the SM between $10^9$ and $10^{15}$~GeV, overlapping the scale where the Higgs quartic vanishes~\cite{Buttazzo:2013uya,Hall:2018let} and the GUT scale.
In terms of Wolfenstein parameters, this relation can be written as
\begin{equation}
    A^2 = (1-\rho)^2 + \eta^2 .
\end{equation}

Ideally we would like to find a UV model that generates this relation without tuning. While we were unable to do so, we investigate an ansatz for the quark Yukawa matrices that can result in such a relation. We find that this ansatz can reproduce the six observed quark masses with five free parameters, once we impose the relation. Yet, it is not clear if that relation is a hint of a UV physics or is just accidental.

%%%%%%%%%%%%%%%%%%%%%%%%%%%%%%%%%%%%%%%%%%%%%%%%%%%%%%%%%%%%%%%%%%%%%%%%%%%%%
\section{CKM matrix running}\label{sec:running}
%%%%%%%%%%%%%%%%%%%%%%%%%%%%%%%%%%%%%%%%%%%%%%%%%%%%%%%%%%%%%%%%%%%%%%%%%%%%%

The CKM elements run due to the fact that the Yukawa couplings run. Furthermore, the running of the CKM matrix is related to the fact that the running of the Yukawa couplings is not universal. If all the Yukawa couplings ran in the same way, the matrices that diagonalize them would not run. Thus, it is the \textit{nonuniversality} of the Yukawa coupling running that results in CKM running. 

Since only the Yukawa coupling of the top quark is large, that is, $\mathcal{O}(1)$, to a good approximation we can neglect all the other Yukawa couplings. There are three consequences of this approximation:
\begin{enumerate}
\item
The CKM matrix elements do not run below $m_t$.
\item
The quark mass ratios are constant except for those that involve $m_t$.
\item
The only Wolfenstein parameter that runs is $A$.
\end{enumerate}
The first two results above are easy to understand, while the third one requires some explanation.
$A$ is the parameter that appears in the mixing of the third generation with the first two generations, and thus is sensitive to the running of the top Yukawa coupling.
 $\lambda$ mainly encodes $1$--$2$ mixing --- that is, between the first and second generations --- and is therefore insensitive to the top quark. The last two parameters, $\eta$ and $\rho$, separate the $1$--$3$ and $2$--$3$ mixing. Thus they are effectively just a $1$--$2$ mixing on top of the $2$--$3$ mixing that is generated by $A$. We see that, to a good approximation, it is only $A$ that connects the third generation to the first and second, and thus it is the only one that runs.

As explained above, there is a disagreement in the literature over the numerical running of the CKM matrix. Ref.~\cite{Balzereit:1998id} works in the limit of vanishing electroweak gauge couplings, arguing that the renormalization group (RG) evolution is dominated by the Higgs sector and strong interactions. Exploiting the large hierarchies between the quark masses and keeping only the mass of the top quark further simplifies the RG equations, to the point that they can be solved analytically. With this method, Ref.~\cite{Balzereit:1998id} reports an increase in $A$ (in the SM) of about $13\%$ from $10^2$~GeV to $10^{15}$~GeV. The authors of Ref.~\cite{Juarezwysozka:2002kx} instead expand the RG equations in terms of the Wolfenstein parameter $\lambda$, noting that the leading term in the one-loop contribution is $\mathcal{O}(1)$ while the leading term in the two-loop contribution is $\mathcal{O}(\lambda^4)$. Motivated by this, they solve the RG equations up to order $\lambda^3$. In Fig.~2 of Ref.~\cite{Juarezwysozka:2002kx} we can see an increase in $A$ of about $25\%$ from $10^2$~GeV to $10^{15}$~GeV, in tension with Ref.~\cite{Balzereit:1998id}.

To address the discrepancy, we compute the running of the Wolfenstein $A$ by directly computing the running of the Yukawa matrices and diagonalizing them to find the Wolfenstein parameters. Following Eqs.~(15)--(22) in Ref.~\cite{Juarezwysozka:2002kx}, the down-type Yukawa matrix at scale $t$
 is given by
\begin{equation}
y_{d}(t)=\sqrt{r^{\prime}(t)} h_{m}^{3}(t)\left(U_{u}\right)_{L}^{\dagger} Z(t)\left(U_{u}\right)_{L} y_{d}^{0},
\end{equation}
where 
\begin{equation} \label{eq:def-t} 
t=\ln\left(\frac{\mu}{\mu_0}\right) .
\end{equation}
$r'(t)$ and $h_m(t)$ are functions and $Z(t)$ is a $3\times 3$ matrix, which we define below, following Ref.~\cite{Juarezwysozka:2002kx}.
Quantities with a zero index are evaluated at the scale $\mu_0$, corresponding to $t = 0$; for example, $y_{d}^{0} \equiv y_{d}\left(t = 0\right)$.
Explicitly, $r'(t)$ is defined in terms of the running gauge couplings by
\begin{equation}
    r'(t) = \prod_{i=1}^{i=3} \left( \frac{ g_i^0 }{g_i(t)} \right)^{2\delta_i}
\end{equation}
where $g_1, g_2$, and  $g_3$ are the $\mathrm{U(1)}_Y$, $\mathrm{SU(2)}_L$, and $\mathrm{SU(3)}_C$ gauge couplings, respectively. The coefficients $\delta_i$ are $\{ 5/82, -27/38, -8/7 \}$ in the SM and $\{ 7/99, 3, -16/9 \}$ in the MSSM\@. $h_m(t)$ is given by
\begin{equation}
    h_m(t) = \exp\left[ \frac{1}{16\pi^2} \int_{0}^{t} Y_t^2(t') dt' \right] ,
\end{equation}
where $Y_t$ is the largest eigenvalue of the up-type Yukawa matrix. (One must not confuse the $t$ defined in Eq.~(\ref{eq:def-t}) with the symbol for the top quark that appears in $Y_t$ and $m_t$.) It is related to the running top quark mass $m_t(t)$ by $\sqrt{2} m_t(t) = v(t) Y_t(t)$, where $v(t)$ is the Higgs VEV\@. Lastly, $Z(t) = \diag \left( 1, 1, h(t) \right)$, where $h(t) = h_m(t)^{-3/2}$ in the SM and $h(t) = h_m(t)$ in the MSSM\@.

The Yukawa matrix $y_d$ can be diagonalized by a scale-dependent bi-unitary transformation
\begin{equation} 
y_d(t)=\left(U_d\right)_L^{\dagger}\tilde y_d(t) \left(U_d\right)_R,
\end{equation} 
where $\tilde y_d$ denotes the diagonal Yukawa matrix, and the $t$-dependencies  of $\left(U_d\right)_L$ and $\left(U_d\right)_R$ are implicit. 
Consider the following hermitian matrix:
\begin{equation}
y_{d}(t) y_{d}(t)^{\dagger} = \left(U_{u}\right)_{L}^{\dagger}\left[r'(t)h_m^6(t)Z(t)\left(U_{u}\right)_{L} y_{d}^{0} y_{d}^{0\dagger}\left(U_{u}\right)_{L}^{\dagger}Z(t)\right]\left(U_{u}\right)_{L} .
\end{equation}
After diagonalizing it we get
\begin{eqnarray}
\tilde y_{d}(t)^2=V(t)^{\dagger}\left[r'(t)h_m^6(t)Z(t)\left(U_{u}\right)_{L} y_{d}^{0} y_{d}^{0\dagger}\left(U_{u}\right)_{L}^{\dagger}Z(t)\right]V(t) ,
\end{eqnarray}
where 
\beq
V(t)\equiv \left(U_u\right)_L(t)\left(U_d\right)_L^\dagger(t)
\eeq
is the CKM matrix at scale $t$.  Consequently, we can directly compute the CKM matrix by diagonalizing  the hermitian matrix 
\begin{equation}
	Z(t)\left(U_{u}\right)_{L} y_{d}^{0} y_{d}^{0\dagger}\left(U_{u}\right)_{L}^{\dagger}Z(t)=\frac{2}{v_0^2}Z(t)V_0(M_d^0)^2V_0^\dagger Z(t),
\end{equation}
where 
$v_0$ is the Higgs vev, $V_0$ is the CKM matrix, and  $M_d^0$ is the mass matrix of down-type quarks (all at the scale $\mu_0$).

Following the definition in the PDG~\cite{Zyla:2020zbs}, we compute the numerical value of Wolfenstein parameters directly from the CKM matrix elements:
\begin{equation}
	\begin{gathered}
	\lambda=\frac{\left|V_{u s}\right|}{\sqrt{\left|V_{u d}\right|^{2}+\left|V_{u s}\right|^{2}}}, \qquad A \lambda=\left|\frac{V_{c b}}{V_{u s}}\right|, \qquad
		A \lambda^{3}(\rho+i \eta)=V_{u b}^{*} 
	\end{gathered} .
\end{equation}
We find that $\lambda$, $\rho$, and $\eta$ only change by $\mathcal{O}(10^{-4})$ from the weak scale to the Planck scale, confirming known results~\cite{Balzereit:1998id,Juarezwysozka:2002kx}. Moreover, the assumption that we use of neglecting all the Yukawa couplings but that of the top is not justified for such a small running. Given the fact that the running of $\lambda$, $\rho$, and $\eta$ is much smaller than that of $A$ we treat them as constants.

\begin{figure}[t]
    \centering
    \includegraphics[width=0.8\textwidth]{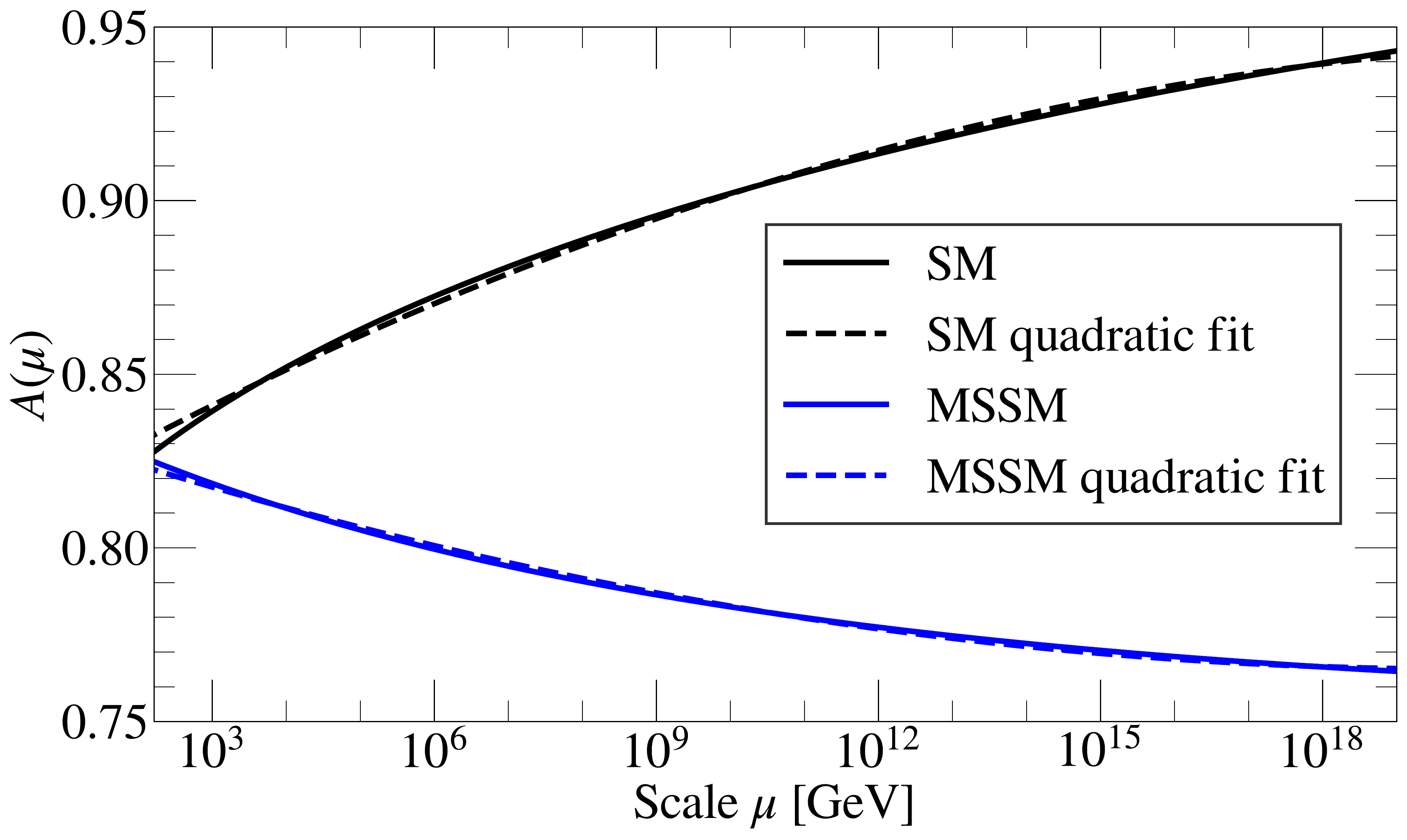}
    \caption{Running of the Wolfenstein parameter $A$ in the SM (black) in the MSSM (blue). In the MSSM we take $\tan \beta = 1$. The running is fitted to the quadratic equations given in Eq.~\eqref{eq:Afit}.}
    \label{fig:running}
\end{figure}

Our results for the running of the Wolfenstein parameter $A$ in the SM and the MSSM are depicted in Fig.~\ref{fig:running}. We use the CKMfitter collaboration central values for the CKM matrix elements~\cite{Charles:2004jd} and the PDG values for all running gauge couplings and masses~\cite{Zyla:2020zbs}. The running of $A$ in the SM agrees with Ref.~\cite{Balzereit:1998id}, contrary to the running reported in Fig.~2 of Ref.~\cite{Juarezwysozka:2002kx}. We conclude that the different methods of computing the CKM running agree with each other, and they are in disagreement with Fig.~2  of Ref.~\cite{Juarezwysozka:2002kx}.

Fitting $A$ to a quadratic equation, we find that it is very well-approximated by
\begin{equation}\label{eq:Afit}\begin{split}
    A^{\text{SM,fit}}(\mu)   &= 0.830 + 0.0111 \log_{10} \left( \frac{\mu}{100\text{ GeV}} \right) - 0.000268 \log_{10}^2 \left( \frac{\mu}{100\text{ GeV}} \right) , \\
    A^{\text{MSSM,fit}}(\mu) &= 0.824 - 0.00653 \log_{10} \left( \frac{\mu}{100\text{ GeV}} \right) + 0.000181 \log_{10}^2 \left( \frac{\mu}{100\text{ GeV}} \right) .
\end{split}\end{equation}
The error of these fits is less than $1\%$ between $m_t$ and $M_{Pl}$.

Lastly, we discuss how a potential CKM relation runs.
We consider some function of CKM matrix elements $R(\mu)$ which, when expressed in terms of the Wolfenstein parameters, scales as $A^n$. Suppose one has computed $R$ at a particular scale $\mu_0$.
Then we have
\begin{equation}
    R(\mu) = \left[ \frac{A(\mu)}{A(\mu_0)} \right]^n R(\mu_0) .
\end{equation}
From this one can see that computing the scale at which $R = 1$ simply amounts to solving the equation
\begin{equation}
    A(\mu) - \frac{ A(\mu_0) }{ R(\mu_0)^{1/n}} = 0.
\end{equation}

%%%%%%%%%%%%%%%%%%%%%%%%%%%%%%%%%%%%%%%%%%%%%%%%%%%%%%%%%%%%%%%%%%%%%%%%%%%%%
\section{CKM relations up to the Planck scale}\label{sec:relations}
%%%%%%%%%%%%%%%%%%%%%%%%%%%%%%%%%%%%%%%%%%%%%%%%%%%%%%%%%%%%%%%%%%%%%%%%%%%%%

\begin{figure}[t!]
    \centering
    \includegraphics[width=0.9\textwidth]{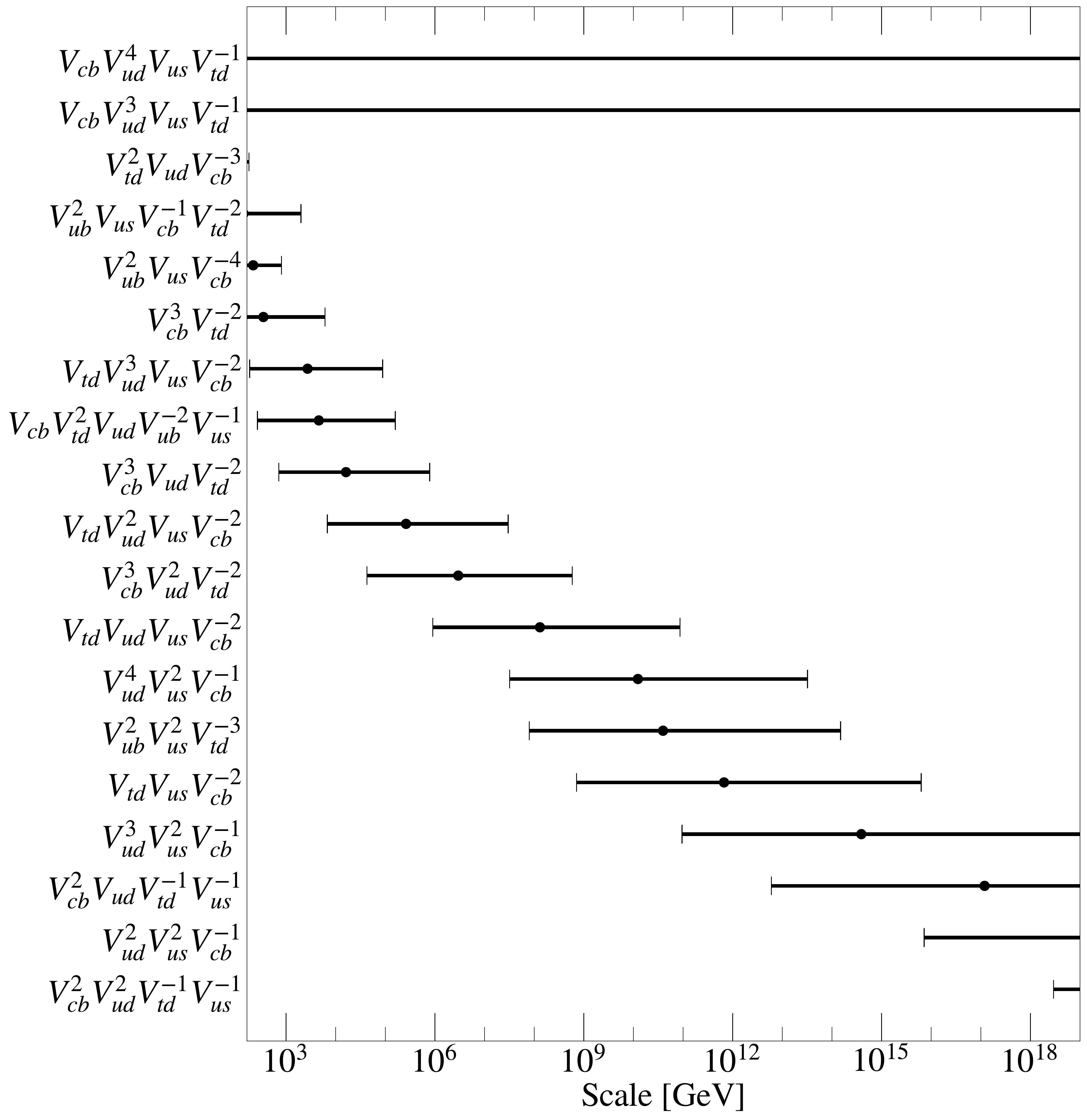}
    \caption{Novel CKM relations in the SM from $m_t$ to $M_{Pl}$. The bars indicate the range of energy scales in which the relations hold to within $2\%$, while the dots indicate the scale at which the relations equal 1. The first two relations do not run at order $\lambda^3$ and so they hold at all scales.}
    \label{fig:relations_SM}
\end{figure}

\begin{figure}
    \centering
    \includegraphics[width=0.9\textwidth]{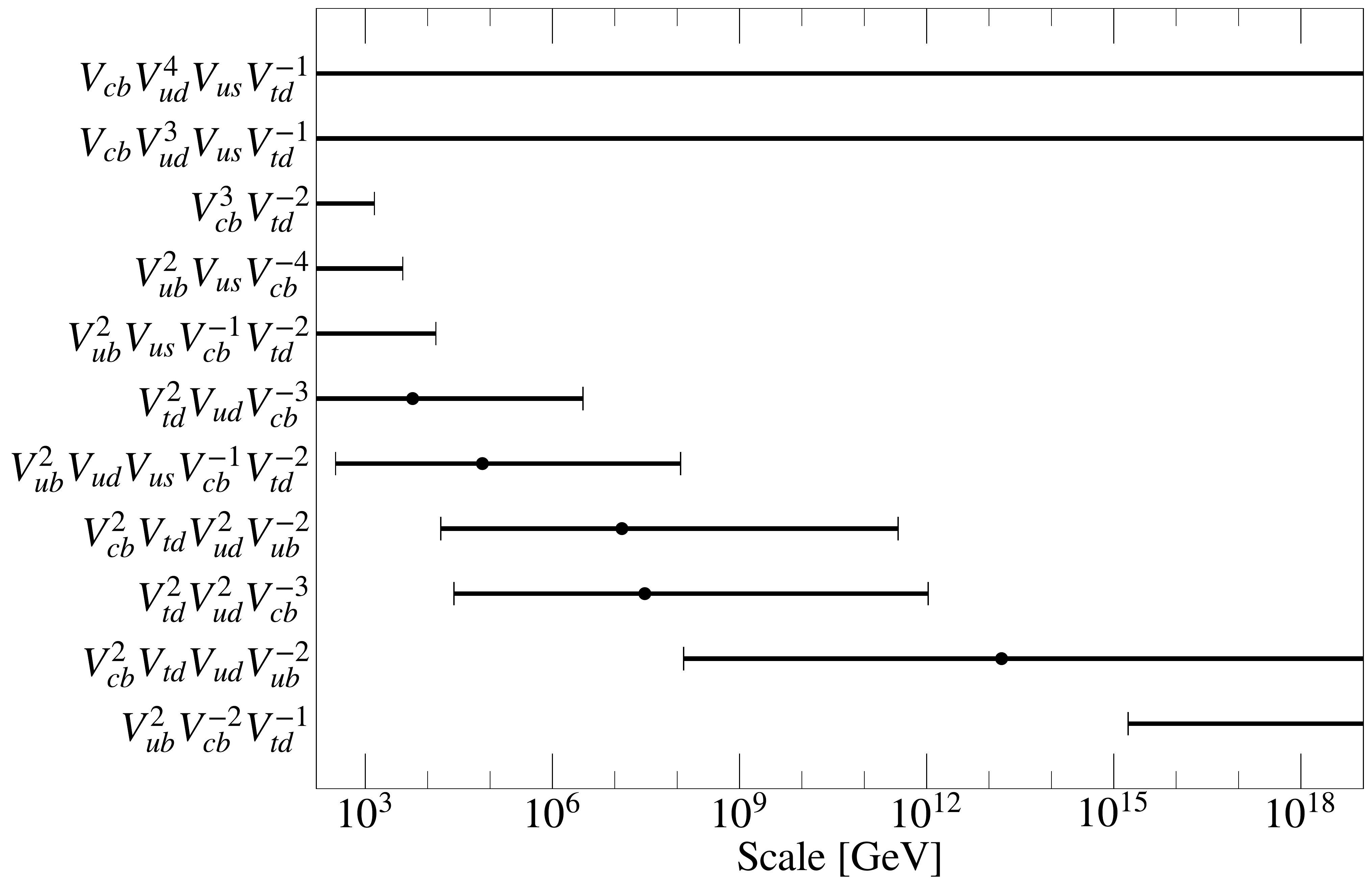}
    \caption{Same as Fig.~\ref{fig:relations_SM}, but for the MSSM\@.}
    \label{fig:relations_MSSM}
\end{figure}

We search for relations across the range of scales $(m_t, M_{Pl})$ of the form
\begin{equation}
    \frac{\prod_n \left| V_{i_n j_n} \right|^{a_n}}{\prod_m \left| V_{k_m l_m} \right|^{b_m}} = 1 \pm 0.02,
\end{equation}
where $i_n, k_m \in (u,c,t)$, $j_n, l_m \in (d,s,b)$, and $\sum_n a_n + \sum_m b_m \leq 7$.
We restrict our attention to those relations which are not already implied by the Wolfenstein parametrization and require that the relations hold to within $2\%$ of 1. We choose a $2\%$ precision because that is roughly the experimental precision on the values of the CKM elements, as well as the theoretical precision in the running formula that we use.
Our results for the SM and MSSM are depicted in Figs.~\ref{fig:relations_SM} and \ref{fig:relations_MSSM}, respectively.

Our analysis differs from Ref.~\cite{Grossman:2020qrp} in two ways. First, Ref.~\cite{Grossman:2020qrp} worked at leading order in $\lambda$, and thus approximated $V_{ud} \approx 1$. We do not make this approximation. Second, we allow $\sum_n a_n + \sum_m b_m \leq 7$, while Ref.~\cite{Grossman:2020qrp} required $\sum_n a_n + \sum_m b_m \leq 6$, i.e.~we allow one more factor of CKM matrix elements.

Two of the relations we find are independent of the Wolfenstein parameter $A$. These are
\begin{equation}
    1 = \frac{|V_{cb} V_{ud}^3 V_{us}|}{|V_{td}|} = \frac{1 - 3\lambda^2/2}{|1 - \rho - i \eta|} + \mathcal{O}(\lambda^4), \qquad 1 = \frac{|V_{cb} V_{ud}^4 V_{us}|}{|V_{td}|} = \frac{1 - 2\lambda^2}{|1 - \rho - i \eta|} + \mathcal{O}(\lambda^4).
\end{equation}
Since these relations are $A$-independent, they do not run.

Of course, at low scales we find the same relations as Ref.~\cite{Grossman:2020qrp}: $\abs{V_{td}^2} = \abs{V_{cb}^3}$ and $\abs{ V_{ub}^2 V_{us} } = \abs{V_{cb}^4}$. In both the SM and the MSSM, these relations hold up to about $10^3$--$10^4$~GeV.

There is a compelling relation in the SM involving only four factors of CKM matrix elements, which holds between $10^9$ and $10^{15}$~GeV:
\begin{equation}\label{eq:relation}
    \abs{V_{td} V_{us}} = \abs{V_{cb}^2} .
\end{equation}
In terms of the Wolfenstein parameters, this relation can be written as 
\begin{equation}
A^2 = \eta^2 + (1 - \rho)^2.
\end{equation}
Below we concentrate on this relation.

%%%%%%%%%%%%%%%%%%%%%%%%%%%%%%%%%%%%%%%%%%%%%%%%%%%%%%%%%%%%%%%%%%%%%%%%%%%%%
\section{Yukawa ansatz}\label{sec:textures}
%%%%%%%%%%%%%%%%%%%%%%%%%%%%%%%%%%%%%%%%%%%%%%%%%%%%%%%%%%%%%%%%%%%%%%%%%%%%%
It would be tantalizing if we could construct a UV model  that generates the relation of Eq.~(\ref{eq:relation}). Yet, we were unable to find one.
Instead, we explored possible forms of the Yukawa matrices that could yield the relation. The idea is that once we find such matrices, they can serve as a first step in constructing a model.

Flavor models, for example Froggatt-Nielsen models~\cite{Froggatt:1978nt,Leurer:1992wg,Leurer:1993gy}, typically predict \textit{textures} for the Yukawa matrices, generating the small Wolfenstein parameter $\lambda$ through some novel dynamics. Although these models explain the hierarchical structure of the CKM matrix, they cannot generate CKM relations (ignoring numerical coincidences), since the Yukawa textures are only defined up to $\mathcal{O}(1)$ parameters. That is, Froggatt-Nielsen models are consistent with Wolfenstein anarchy.

For a flavor model to generate a relation like Eq.~\eqref{eq:relation}, it must predict a precise form for the Yukawa matrices, without unspecified $\mathcal{O}(1)$ constants.
In fact, we are unaware of any model that can generate such a relation.

There are ten observables in the SM quark sector (six masses, three mixing angles, and one complex phase). Thus, an ansatz for the Yukawa matrices which correctly predicts all of these observables is only nontrivial if it involves fewer than ten input parameters. But even this is quite an ambitious goal. Here we achieve a more pedestrian goal: we find an ansatz that reproduces the correct quark masses with fewer than six parameters, while making a connection to the relation in Eq.~\eqref{eq:relation}.

Consider the following quark mass matrix ansatz:
\begin{equation}\label{eq:texture}
    M_u = m_t \begin{pmatrix} 0 & x \lambda_u^3 & 0  \\ x \lambda_u^3 & x \lambda_u^2 & x \lambda_u \\ 0 & x \lambda_u & 1 \end{pmatrix} , \qquad
    M_d = m_b \begin{pmatrix} 0 & w \lambda_d^3 & 0  \\ w \lambda_d^3 & w \lambda_d^2 & 0 \\ 0 & 0 & 1 \end{pmatrix} .
\end{equation}
This ansatz is inspired by the phenomenologically relevant textures identified in Refs.~\cite{Ramond:1993kv,Fritzsch:1999ee}. There are six free parameters $m_t, m_b, x, w, \lambda_u$, and $\lambda_d$. Note that we have made two simplifications.  First, we take all parameters real (and therefore neglect the CKM phase).  Second, while normally each entry is multiplied by an order-one free parameter, and therefore this texture would include 10 free parameters (after accounting for texture zeroes), we have reduced the 10 parameters to 6 parameters.  These simplifications mean that the following toy analysis cannot fully describe nature.  In what follows we treat $\lambda_u$ and $\lambda_d$ as small, as they tend to be on the order of the Wolfenstein parameter $\lambda \approx 0.22$. As we show below, if we require additionally that the relation in Eq.~\eqref{eq:relation} is satisfied, the number of parameters is reduced to five. We further show that one can choose parameters that reproduce the six quark masses observed in nature while satisfying the relation in Eq.~\eqref{eq:relation}.  Note that we do not attempt to explain the precise size of the CKM mixing angles in this toy analysis.

The eigenvalues of $M_u$ are given by
\begin{equation}
    m_t , \qquad m_c \simeq x (1 - x) \lambda_u^2 m_t , \qquad m_u \simeq x (1 - x)^{-1} \lambda_u^4 m_t,
\end{equation}
where we are retaining terms at the leading order in $\lambda_u$. Similarly the eigenvalues of $M_d$ at the leading order in $\lambda_d$ are
\begin{equation}
    m_b , \qquad m_s \simeq w \lambda_d^2 m_b , \qquad m_d \simeq w \lambda_d^4 m_b .
\end{equation}
While we do not write them explicitly, we note that the eigenvectors of $M_d$ and $M_u$ are independent of $m_b$, $m_t$, and $w$. For $m_t$ and $m_b$ this is easy to see, since they are overall factors multiplying the matrices. To see that this is also the case for $w$, observe that $M_d$ is block diagonal and $w$ is an overall factor multiplying the upper block.
Thus, we conclude that the CKM matrix depends only on the parameters $\lambda_u$, $\lambda_d$, and $x$.

\begin{figure}[t]
    \centering
    \includegraphics[width=\textwidth]{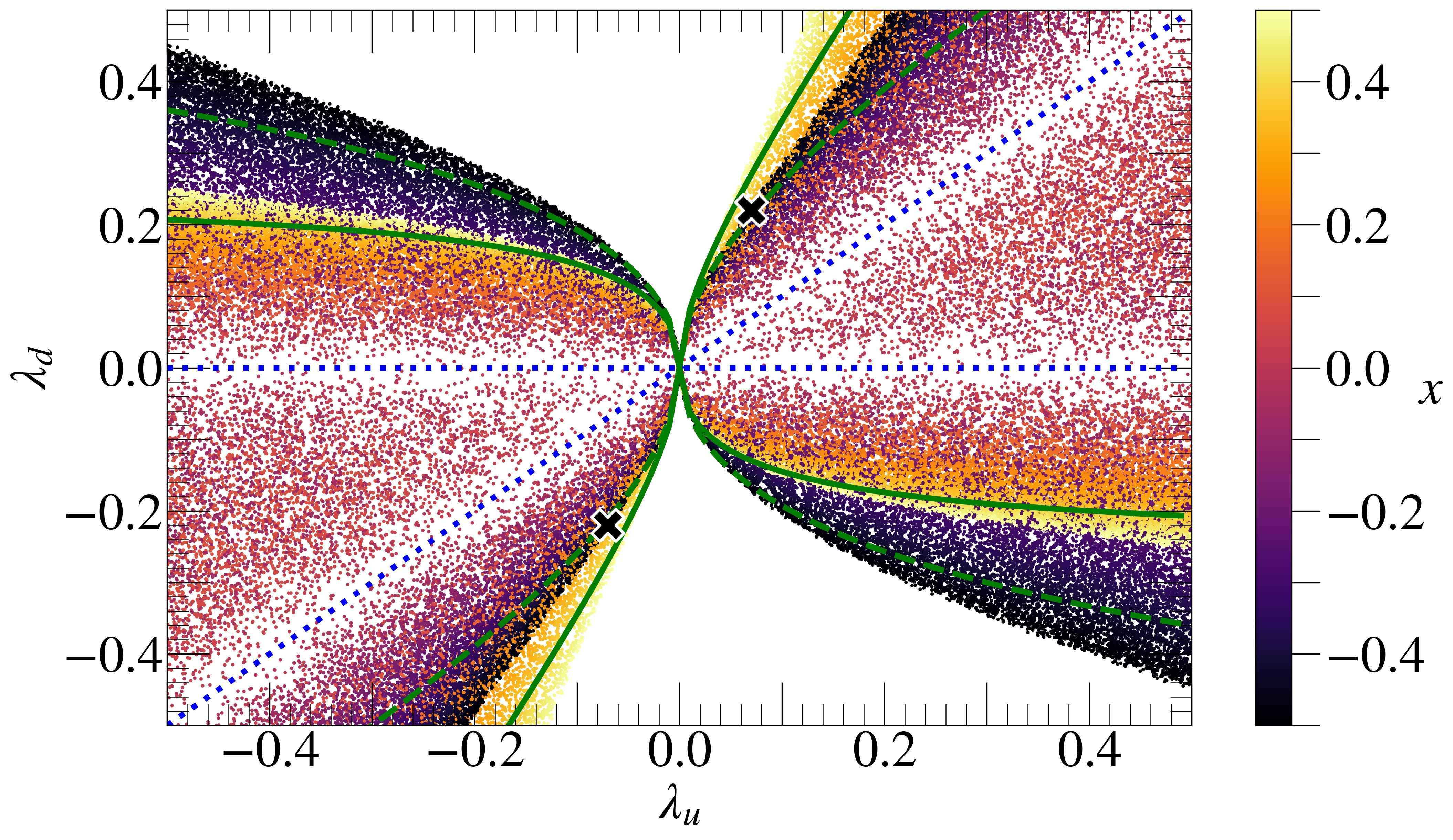}
    \caption{Parameters of random mass matrices generated from the ansatz in Eq.~\eqref{eq:texture} which lead to a CKM matrix satisfying the relation~\eqref{eq:relation}. Black X marks indicate values of $\lambda_u$, $\lambda_d$ which reproduce the observed quark mass hierachies with $x \approx -0.5$. Green lines correspond to the analytical estimate in Eq.~\eqref{eq:firstordersolution} for $x = +0.5$ (solid line) and $x= -0.5$ (dashed line). Blue dotted lines indicate the lines $\lambda_d = 0$ and $\lambda_d = \lambda_u$.}
    \label{fig:scatter}
\end{figure}

We generate $10^7$ random Yukawa matrices from this ansatz, uniformly sampling the parameters $\lambda_d$, $\lambda_u$, and $x$ from the range $[-0.5, 0.5]$. About $7.5 \times 10^4$ of these satisfy the relation in Eq.~\eqref{eq:relation} to within 2\%.  (One might guess that about 2\% of matrices would satisfy the relation within 2\%; this turns out to be correct up to an order one factor.)
The parameter values giving rise to matrices satisfying the relation are depicted in Fig.~\ref{fig:scatter}.

To better understand the features of this plot, one can solve for the CKM matrix perturbatively in $\lambda_u$ and $\lambda_d$, then extract the ratio 
\begin{equation}
R \equiv \left|\frac{V_{td} V_{us}}{V_{cb}^2}\right|.
\end{equation}
The relation in Eq.~\eqref{eq:relation} corresponds to $R = 1$. 
Up to order $\lambda^4$, we find
\begin{equation}\label{eq:approximation}\begin{split}
    R &= \left | \frac{-\lambda_u \lambda_d + \lambda_d^2 (1-x) }{\lambda_u x (1-x)} \right . \\
    &+ \frac{1}{2 x \lambda_u (1 - x)^3 } \left[ 2 \lambda_u ^4 (x-1)^2 x - \lambda_u^3 \lambda_d \left((x-1)^2 x^2-1\right) \right. \\
    & \left . \left. - \lambda_u^2 \lambda_d^2 (x-1) \left(x (3 x-2) (x-1)^2+1\right) + 4 \lambda_u \lambda_d^3   (x-1)^2 + 6 \lambda_d^4 (x-1)^3 \right] \vphantom{\frac{-\lambda_u \lambda_d + \lambda_d^2 (1-x) }{\lambda_u x (1-x)}} \right | \\
    &+ \mathcal{O}(\lambda^5) .
\end{split}\end{equation}
In the above calculation we have assumed that $\abs{x} < 1$.
Note that $R$ is unchanged under the transformation $\lambda_u \rightarrow - \lambda_u$, $\lambda_d \rightarrow - \lambda_d$ (while holding $x$ constant). This is the reason for the symmetry observed in Fig.~\ref{fig:scatter}.

If we retain only the leading-order, $\mathcal{O}(\lambda^2)$ term, it is easy to solve for $\lambda_d$ as a function of $\lambda_u$ and $x$:
\begin{equation}\label{eq:firstordersolution}
    \lambda_d \simeq \frac{1}{2} x \lambda_u \left[ \frac{1}{x(1-x)} \pm \sqrt{\frac{1}{x^2(1-x)^2} + \frac{4}{x \lambda_u} } \,\right] .
\end{equation}
In accordance with the discussion above, there is another solution obtained by taking $\lambda_u \rightarrow - \lambda_u$ and $\lambda_d \rightarrow -\lambda_d$. Setting $x = \pm 1/2$ roughly gives the boundaries of the region of parameter space populated by our scan, at least at small values of $\lambda_u$ and $\lambda_d$ (see the green lines in Fig.~\ref{fig:scatter}).

Eq.~\eqref{eq:firstordersolution} exemplifies our previous statement that imposing the relation Eq.~\eqref{eq:relation} upon our ansatz effectively reduces the number of free parameters from six to five. 
Next we show that it is possible to choose the parameters so that the relation is satisfied and all six quark masses are reproduced correctly.
We first choose the parameters $\lambda_u$, $\lambda_d$, $x$, and $w$ to reproduce the observed quark mass ratios \cite{Zyla:2020zbs,Xing:2007fb}. For the mass ratios at the GUT scale we use
\begin{equation}
{m_u \over m_c} \approx 0.002, \qquad
{m_c \over m_t} \approx 0.003, \qquad 
{m_d \over m_s} \approx 0.05, \qquad 
{m_s \over m_b} \approx 0.02. 
\end{equation}
Then, the overall constants in front of $M_u$ and $M_d$ can be chosen to reproduce the correct running values of $m_t$ and $m_b$.

Specifically, one takes $\lambda_u = -0.07$ and $x = -0.50$, which reproduces the observed ratios $m_u/m_c$ and $m_c/m_t$. Imposing the relation~\eqref{eq:relation} then requires $\lambda_d \approx -0.22$ --- which also gives the correct mass ratio $m_d/m_s$. Then, one can choose $w \approx 0.36$ so as to reproduce the correct value for $m_s/m_b$. These values of $\lambda_u$ and $\lambda_d$ are indicated on Fig.~\ref{fig:scatter} by black X marks.

One would expect that a Yukawa ansatz with six free parameters would not be able to simultaneously satisfy Eq.~\eqref{eq:relation} and correctly yield all six quark masses. At best, a generic Yukawa ansatz with six parameters could either satisfy Eq.~\eqref{eq:relation} and give five correct quark masses, or violate Eq.~\eqref{eq:relation} and get all quark masses correct. In this sense, the ansatz in Eq.~\eqref{eq:texture} is not generic. This is reflected in that the choice of $\lambda_d \approx -0.22$ in the preceding paragraph, which was needed to satisfy the relation Eq.~\eqref{eq:relation}, also happened to yield the correct value for $m_d/m_s$.

Next we  consider the regime $x \ll 1$, which is relevant for the interior region of the plot. Expanding Eq.~\eqref{eq:approximation} at leading order in $x$ yields
\begin{equation}
    R \simeq \frac{1}{x} \left[P(\lambda_u, \lambda_d)+ \mathcal{O}(x)\right],
\end{equation}
where $P$ is the polynomial
\begin{equation}
    P(\lambda_u, \lambda_d) = -\lambda_d + \frac{\lambda_d^2}{\lambda_u} + 2 \lambda_d^3 - \frac{3\lambda_d^4}{\lambda_u} + \frac{1}{2} (\lambda_d \lambda_u^2 + \lambda_d^2 \lambda_u ) .
\end{equation}
There are three solutions to $P = 0$:
\begin{equation}\label{eq:zerosolutions}
    \lambda_d = 0 , \qquad \lambda_d = \lambda_u , \qquad \lambda_d = \frac{1}{6} \left( - \lambda_u \pm \sqrt{12 - 5 \lambda_u^2} \right) .
\end{equation}
When $P = 0$, it is impossible to satisfy $R = 1$.
Hence, the first two solutions in Eq.~\eqref{eq:zerosolutions} correspond to the gaps in Fig.~\ref{fig:scatter} (dotted blue lines). The last solution lies outside of the plot range.
As one moves slightly away from these lines, one can tune $x = P(\lambda_u, \lambda_d)$ to satisfy the relation. 

The above toy analysis illustrates some of the features that a UV explanation of $R=1$ should have.   Even after taking the simplified texture of Eq.~(\ref{eq:texture}), with 10 nonzero complex parameters simplified to 6 real parameters, we find that the texture parameters should satisfy the nontrivial relation of Eq.~(\ref{eq:firstordersolution}), in order to enforce $R=1$.  Ideally, $R=1$ would be a structural consequence of a UV model, and not a parametric accident.  This seems challenging to realize.

%%%%%%%%%%%%%%%%%%%%%%%%%%%%%%%%%%%%%%%%%%%%%%%%%%%%%%%%%%%%%%%%%%%%%%%%%%%%%
\section{Conclusions}\label{sec:conc}
%%%%%%%%%%%%%%%%%%%%%%%%%%%%%%%%%%%%%%%%%%%%%%%%%%%%%%%%%%%%%%%%%%%%%%%%%%%%%

In this work, we have explored novel CKM relations that are Wolfenstein-independent, in the sense that they are not implied purely by the smallness of $\lambda$ in the Wolfenstein parametrization. We looked for them from the weak scale up to the Planck scale, computing the running in the SM and one realization of the MSSM\@. This builds upon previous work in Ref.~\cite{Grossman:2020qrp} which examined such relations at low scales. In particular, the relation in Eq.~\eqref{eq:relation}, which holds in the SM near the GUT scale, is rather simple.

We have also settled the disagreement between Ref.~\cite{Balzereit:1998id} and Ref.~\cite{Juarezwysozka:2002kx} over the running of the CKM matrix. Although Ref.~\cite{Juarezwysozka:2002kx} reports a different running of $A$ from Ref.~\cite{Balzereit:1998id}, recomputing the running using their methods yields a result that agrees with Ref.~\cite{Balzereit:1998id}. Furthermore, our results confirm that to a very good approximation, the only Wolfenstein parameter that runs is $A$ --- the other parameters are effectively constant. We provide quadratic fits to the running of $A$ in the SM and the MSSM in Eq.~\eqref{eq:Afit}, which are valid to within $1\%$ between $m_t$ and $M_{Pl}$. These results may be of use to others interested in the CKM matrix at different scales.

It seems difficult to construct UV models that explain the sort of CKM relations we consider. Flavor models usually only constrain the Yukawa matrices up to $\mathcal{O}(1)$ factors, and so they do not make any precise predictions of CKM relations. It would be quite interesting and challenging to find a UV model that can dynamically generate CKM relations like Eq.~\eqref{eq:relation}.

Indeed, we were unable to find a UV model to explain Eq.~\eqref{eq:relation}. We instead investigated an ansatz for the Yukawa matrices, Eq.~\eqref{eq:texture}. If one imposes the relation Eq.~\eqref{eq:relation} upon the ansatz, there are effectively five free parameters. Interestingly, this ansatz can still correctly reproduce all six quark masses.

The big question is what the implications of the relations we found are. It would be nice if they will lead us into any UV physics. Yet, at this point we do not see any, and the relations may be just accidental. 

%%%%%%%%%%%%%%%%%%%%%%%%%%%%%%%%%%%%%%%%%%%%%%%%%%%%%%%%%%%%%%%%%%%%%%%%%%%%%
\section*{Acknowledgements}
%%%%%%%%%%%%%%%%%%%%%%%%%%%%%%%%%%%%%%%%%%%%%%%%%%%%%%%%%%%%%%%%%%%%%%%%%%%%%

The work of YG is supported in part by the NSF grant PHY1316222.
AI is supported by NSERC (reference number 557763) and by NSF grant PHY-2014071. JTR is supported by the NSF CAREER grant PHY-1554858 and NSF grant PHY-19154099. TT is supported by the MoST (MoST-109-2917-I-007-001).

\bibliography{References}
\end{document}